\documentclass{ws-ijmpcs}
\usepackage{graphics}

\DeclareGraphicsExtensions{.jpg, .png , .pdf}

\begin{document}

\markboth{Noemie Globus \& Amir Levinson}
{Plasma injection and outflow formation in Kerr black holes}

%
\catchline{}{}{}{}{}
%

\title{Plasma injection  and outflow formation in Kerr black holes}

\author{Noemie Globus and Amir Levinson}
\address{School of Physics \& Astronomy, Tel Aviv University Tel Aviv 69978, Israel}

\maketitle

\begin{history}
\received{Day Month Year}
\revised{Day Month Year}
\end{history}

\begin{abstract}
We discuss the role plasma injection plays in the formation of outflows in Kerr spacetime. Using a model for the double flow established in the polar region of a rotating black hole, we study the interplay between the different processes that can power the outflow. 
In particular, we find two types of flows with distinct properties that depend on the rate at which energy is deposited in the magnetosphere. We discuss the implications of this result for gamma ray bursts outflows.
\end{abstract}

\ccode{PACS numbers: 04.70.-s, 47.75.+f, 95.30.Qd}

\section{Introduction}

Accreting black holes are thought to power the relativistic jets that form in active galactic nuclei (AGNs), microquasars, and gamma-ray bursts (GRBs).  A plausible production mechanism for those jets is magnetic extraction of the spin energy of a Kerr black hole\cite{BZ77}.   A key feature of this process is a double trans-magnetosonic plasma flow which is launched from a stagnation radius located between the inner and outer light surfaces, and sustained by a plasma source in the magnetosphere.  In general, there is
a range of plasma injection rates within which the Blandford-Znajek (BZ) process can be activated;  it has to be sufficiently high to provide the charge density required to establish an MHD flow, but low enough to avoid overloading of magnetic field lines that leads to a shutdown of the BZ process.  

The nature of the plasma source in most of the relativistic systems mentioned above is not well understood yet.   In AGNs the problem seems to be how to inject enough charged particles on open magnetic field lines.    Direct feeding by the surrounding accretion flow seems unlikely, as charged particles would have to cross magnetic field lines on timescale shorter than the accretion time in order to reach the polar outflow.  Free neutrons that may be produced in a radiative inefficient accretion flow (RIAF) can cross field lines, however, they will decay over a distance typically much shorter than the horizon scale of a supermassive black hole.  Thus, even if existent at sufficient quantity they will not reach the inner regions.    Pair production on open filed lines via annihilation of MeV photons is a plausible plasma source, but requires sufficiently hot accretion flow, and may be relevant only to faint sources like M87 and  Sgr $A^*$.  
It has been shown that the density of the charges thereby injected depends sensitively on the accretion 
rate and the conditions in the RIAF \cite{LR11,MGDS11} .  Naive estimates\cite{LR11}, although highly uncertain, suggest that in case of M87 this process cannot provide complete screening at accretion rates that correspond to the inferred jet power, and to a fit of the observed SED  by an advection-dominated accretion flow (ADAF)  model.  As a consequence, an intermittent gap may form at the base of the flow in which the injected density is amplified to the required level by copious pair cascades induced by the potential drop in the gap.  This gap may be the source of the variable TeV emission observed in M87 and, conceivably, some other non-blazar AGNs\cite{Le00,NA07,LR11}. 

In powerful blazars the accretion flow is much cooler, hence direct charge injection should not ensue. However, the disk luminosity is much higher and it could be that once pair creation is initiated in the gap by a stray charge, it will be sustained forever in a cyclic process,  owing to inverse Compton scattering and pair creation on the dense target photon field, as demonstrated in the case of pulsars\cite{Tmk10,TA13}.  

 In GRBs the plasma injection rate is anticipated to be always well above that required to establish a MHD flow.  However, under certain conditions, it may lead to overloading and a consequent shutdown of the BZ process.   In what follows, we describe some recent studies\cite{LG13,GL13} wherein the role of plasma injection in the magnetosphere of a Kerr black hole has been carefully examined.   
 \section{A model for loaded MHD flows}
  
We consider relativistic jets that are created near the polar region of a Kerr black hole.  As mentioned above, the magnetosphere consists of a double-flow structure, whereby two plasma streams are launched, along every magnetic flux tube, in opposite directions from a stagnation radius $r_\textrm{st}(\theta)$ located between the two light surfaces, where $\theta$ is the inclination angle of the flux tube.  To analyze this structure, we have constructed\cite{GL13} a semi-analytic model that incorporates plasma injection in a self-consistent manner. The injection process is modeled by prescribed source terms in the MHD equations, that determine the rate of change in mass, energy, angular momentum and entropy along streamlines. This generalize the equations of Ref. \refcite{Lev06b} to the Kerr geometry.

\begin{figure}[pb]
\centerline{\psfig{file=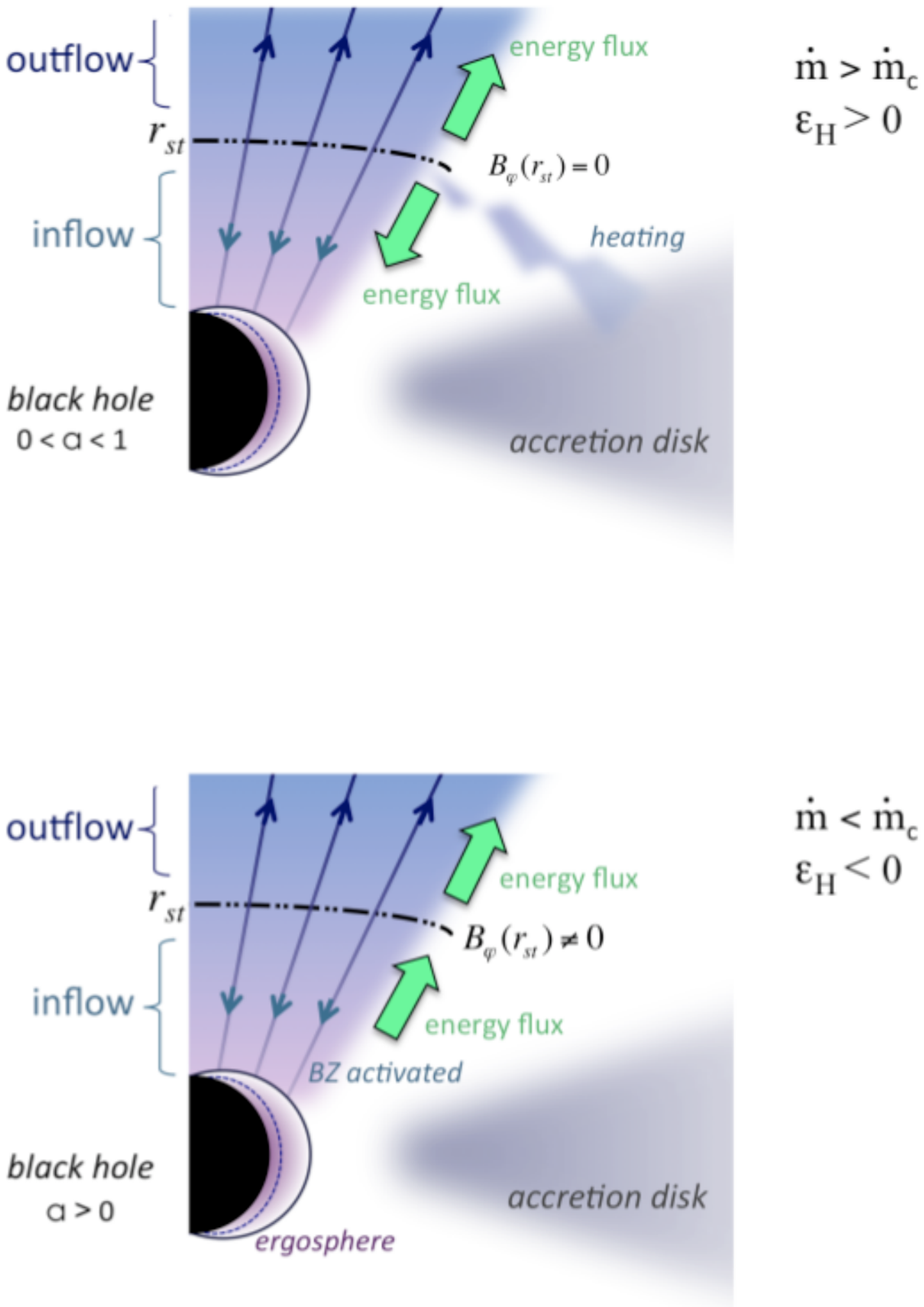,width=17cm}}
\vspace*{8pt}
\caption{Illustration of the double-flow structure: In a Kerr spacetime there exist two types of flows, depending on the plasma injection rate. At supercritical loads the outflow is powered by the external energy source (upper panel). At subcritical loads, the outflow is powered by the black hole rotational energy (lower panel).\label{f1}}
\end{figure}

We identified two distinct types of solutions, that correspond to regimes where the BZ process is switch-on or switch-off (see Figure \ref{f1}).   These two types of flows are characterized by the sign of the specific energy on the horizon, henceforth denoted by ${\cal E}_H$.
For negative energy solutions (${\cal E}_\textrm{H}<0$) the energy flux is always directed outwards, implying energy extraction of the black hole spin energy.   At sufficiently low injection rates the emitted power is shown to converge to the BZ result derived in the force free limit (see below).
The dynamics of the flow in this regime is governed entirely by the frame dragging potential induced by the black hole. 
In case of positive energy solutions (${\cal E}_\textrm{H}>0$) the dynamics of the flow is dictated by the external plasma source. If the plasma injected in the magnetosphere is relativistically hot, then a pressure driven, double-flow is launched from the stagnation radius, whereby a fraction of the injected energy is absorbed by the black hole and the rest emerges at infinity in the form of a relativistic jet.  If the injected plasma is cold, an outflow may not form at all. 

Which type of flow will form under given conditions depends merely on the plasma injection rate, as discussed in the next section. 

\section{A critical load}	
Takahashi et al.\cite{TNTT90} have shown that two conditions must be satisfied in order for energy to be extracted from a Kerr  black hole: (i) the frame dragging potential must exceed the angular velocity of the magnetic field lines near the horizon, and (ii) the Alfv\'en point of the inflow must be located inside the ergosphere. When these conditions are satisfied, the specific energy of a fluid element near the horizon, as measured at infinity, is negative, ${\cal E}_H<0$, implying an outward energy flux on the horizon. 

A question of interest is how the efficiency of the extraction process depends on the load.  This question was not addressed in Ref. \refcite{TNTT90}.  In order to analyze the effect of the load, we computed the structure of the ideal MHD inflow emanating from the stagnation radius, for different plasma injection rates.  To simplify the analysis, we invoked an infinitely thin injection zone, whereby the mass injection profile is given by ${q_n}\propto \delta(r-r_{st})$,  and likewise for the energy-momentum source terms.  Then, the specific energy ${\cal E}$ is conserved at $r<r_{st}$, and its value is uniquely determined by the regularity condition at the fast magnetosonic point.  The value of the enthalpy flux injected at the stagnation radius fixes the location of the slow point\footnote{in the cold case, the slow point coincides with the stagnation point}.  For further details see Ref. \refcite{GL13}. 

\begin{figure}[!h]
{\psfig{file=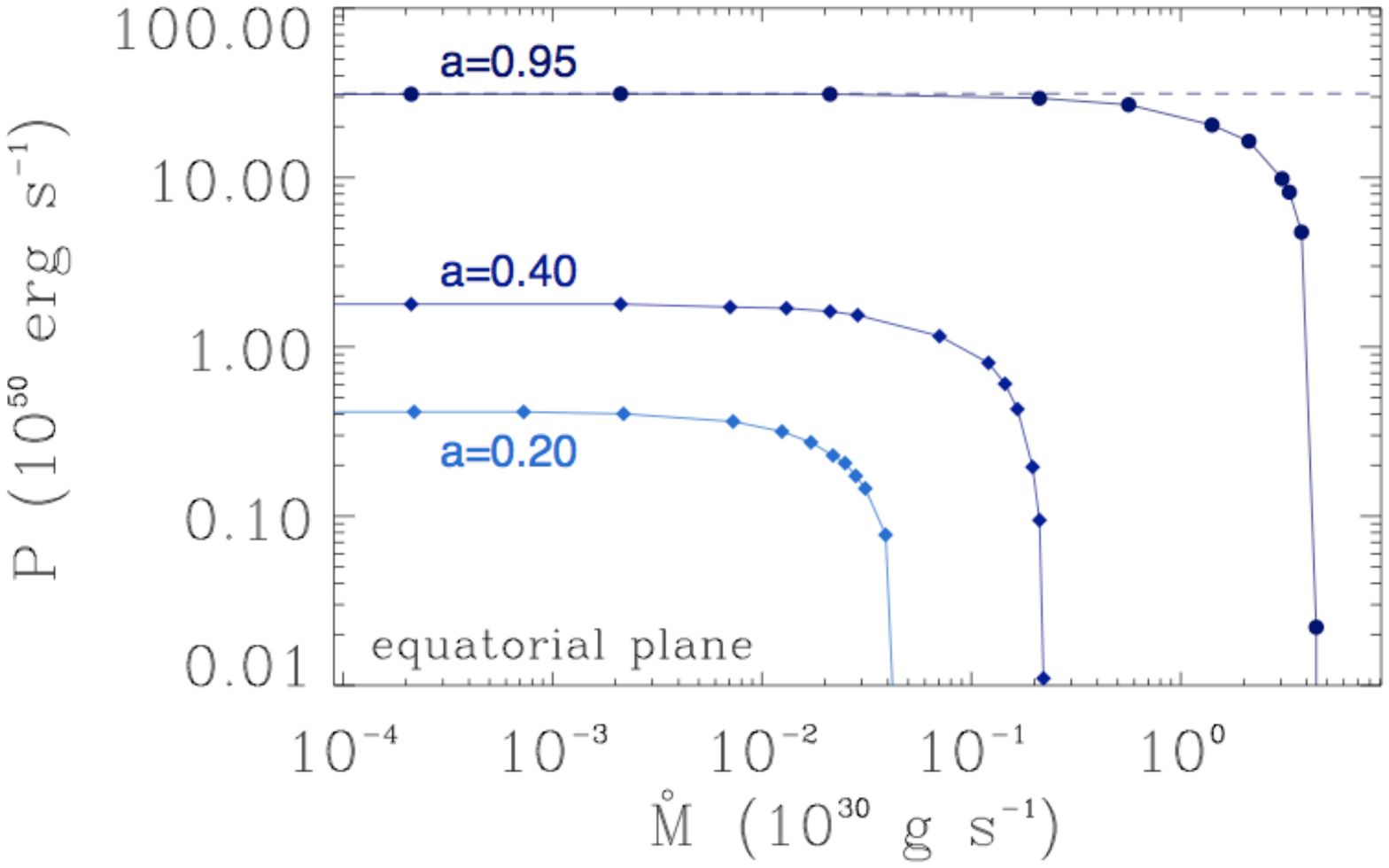,width=14cm}}
\vspace*{8pt}
\caption{Extracted power of an equatorial flow ($\theta=\pi/2$) vs. the injected mass flow rate in the regime where energy extraction is switched on, for different black hole angular momenta $a$. Each point corresponds to a cold inflow solution. The horizontal dashed line gives the force-free result derived in Ref. 1. The critical load $\dot{M}_c$ can be readily obtained from the figure in each case. \label{f2}}
\end{figure}

To elucidate key features, we consider first the zero temperature limit.   An example is shown in Figure \ref{f2} for an equatorial flow, where the extracted power $P$ is plotted against the injected mass flow rate $\dot{M}$.  The numerical values  were computed assuming magnetic flux of $\Psi=9\times10^{27}$ G cm$^2$.   As seen, the extracted power converges to the force-free result, $P_{FF}$, derived in Ref. \refcite{BZ77} (marked by the horizontal dashed line) at sufficiently small loads, but is strongly suppressed as the load approaches the critical value $\dot{M}_c\simeq P_{FF}/c^2$.
From this, we concluded that a rotating black hole can transfer its rotational energy to the outflow only along field lines on which the 
accretion rate satisfies 
\begin{equation}
\dot{M}< 10^{-4}\left(\frac{M_{BH}}{3M_{\odot}}\right)^{-2}\left(\frac{\Psi_0}{10^{27}{\rm G\,cm^2}}\right)^{2}g(a,\theta)\quad {\rm M_{\odot}\, s^{-1}},
\label{activation-cond}
\end{equation}
where $g(a,\theta)=a^2(r_H^2+a^2)\sin^2\theta/[r_H^2(r_H^2+a^2\cos^2\theta)]$.  

In Ref. \refcite{GL13} it is shown that similar results are obtained in the general case of a hot flow.  
The critical condition generalizes to $P_{inj}\simeq P_{FF}/c^2$, where $P_{inj}$ is total power injected in the magnetosphere which, in the zero temperature limit, reduces to $P_{inj}=\dot{M}c^2$.

\section{Application to GRBs}	

The immediate consequence of the above results for long GRBs is that following the stellar collapse, the polar region must be devoid of matter
in order for a jet to form.  But even then, there is another plasma source in the magnetosphere, namely annihilation of MeV neutrinos that emanate from the hyper-accretion disk surrounding the black hole.   The plasma thereby deposited is relativistically hot, and so a polar outflow will be driven either by the black hole or by the pressure of the injected plasma, provided that the central region is baryon poor, as explained above.

MeV neutrinos are emitted from the inner disk region, with a sensitive dependence of the neutrino luminosity on accretion rate. 
Recent calculations of the annihilation rate around a Kerr black hole\cite{CB07,ZB11} yield a net energy deposition rate of $\dot{E}_{\nu\bar{\nu}}\simeq 10^{52} \dot{m}_\textrm{acc}^{9/4}\left({M_\textrm{BH}}/{3 M_{\odot}}\right)^{-3/2}x_\textrm{mso}^{-4.8}$ erg s$^{-1}$, for accretion rates (henceforth measured in units of $M_{\odot}$ s$^{-1}$) in the range $0.02<\dot{m}_\textrm{acc}<1$, where $x_\textrm{mso}$ is the radius of the marginally stable orbit in units of $m=GM_\textrm{BH}/c^2$.  From our analysis we estimate that for accretion rates
\begin{equation}
\dot{m}_{acc}<0.1\left(\frac{M_{BH}}{3M_\odot}\right)^{-2/9}\left(\frac{\Psi_0}{10^{27}{\rm G\ cm^2}}\right)^{8/9}f(a,\theta)\label{active-BZ}\,,
\end{equation}
the jet is powered by the black hole (the first flow type), whereas for higher rates it is powered by the neutrino source.
The  function $f(a,\theta)$ satisfies $f(0,\theta)=0$, but otherwise depends weakly on $a$.  For $\theta=\pi/2$ it varies between 1 and $1.2$ in the range $0.95\ge a\ge 0.2$.

In a future work, we will intend to generalize the analysis of Ref. \refcite{GL13} to realistic injection profiles, as those computed in Ref. \refcite{ZB11}.  In a preliminary work\cite{LG13}, 
we already analyzed the structure of a neutrino driven flow in a Schwarzschild geometry. We restricted our analysis to unmagnetized and non rotating flows, since the hole itself is non rotating ($a=0$) and the particles follow radial geodesics near the horizon. We adopted an energy deposition profile of the form ${-q_t}\propto x^{-b}$  with $b\simeq4.5$ for $a=0.95$ and  $b\simeq3.5$ for $a=0$, where $a$ is the normalized black hole spin\cite{ZB11}. We derived the double transonic structure and computed the power of the outflow for different energy deposition profiles\cite{LG13}. We requiered that both the inflow and the outflow solutions pass smootly through their sonic points. We started the integration at the inner sonic point and adjusted the stagnation pressure to cross the outer one. We found that for a given choice of the energy deposition profile, there exists a unique solution that passes through the inner and the outer sonic points. Near the horizon, the inflow moves along radial geodesics, while the position of the outer sonic point is determined by the pressure at the stagnation point. 

We concluded that the outflow production efficiency $\epsilon$, defined as the fraction of $\dot{E}_{\nu\bar{\nu}}$ that emerges at infinity, is typically large,  with $\epsilon=0.58$, $0.73$, $0.83$ for $b=5$, $4$ and $3.5$, respectively.  We also found that the specific entropy in the outflow is larger than usually thought (see Figure 2 of Ref.~\refcite{LG13}), owing to the delayed acceleration of the flow, and pointed out the implications for prompt emission.



\begin{thebibliography}{0}    
\bibitem{BZ77} R. D. Blandford and R. L. Znajek, {\it Mon. Not. R. Astron. Soc.}  {\bf 179}, 433 (1977).
\bibitem{LR11} A. Levinson and F. Rieger,  {\it Astrophys. J.} {\bf 730}, 123 (2011).
\bibitem{MGDS11} M. Mo\'scibrodzka, C. F. Gammie,  J. C. Dolence and H. Shiokawa, {\it Astrophys. J.} {\bf 735}, 9 (2011). 
\bibitem{Le00} A. Levinson,  {\it Phys. Rev. Lett.},  {\bf 85}, 912 (2000).
\bibitem{NA07} A. Neronov and  F. A., Aharonian,  {\it Astrophys. J.} {\bf  671}, 85 (2007).
\bibitem{Tmk10} A. Timokhin, {\it Mon. Not. R. Astron. Soc.} {\bf  408}, 209 (2010).
\bibitem{TA13} A. Timokhin and J. Arons, {\it Mon. Not. R. Astron. Soc.} {\bf  429}, 429 (2013).
\bibitem{LG13} A. Levinson and N. Globus, {\it Astrophys. J.}  \textbf{770}, 159 (2013).
\bibitem{GL13} N. Globus and A. Levinson, Loaded MHD flows in Kerr spacetime, to appear in {\it Phys. Rev. D15}. 
\bibitem{Lev06b} A. Levinson, {\it Astrophys. J.} \textbf{648}, 510 (2006).
\bibitem{TNTT90} M. Takahashi, S. Nitta, Y. Tatematsu and A. Tomimatsu, {\it Astrophys. J.} {\bf 363}, 206 (1990).
\bibitem{CB07} W.-X. Chen and A. Beloborodov, {\it Astrophys. J.}  \textbf{657}, 383 (2007).
\bibitem{ZB11} I. Zalamea and A. Beloborodov, {\it Mon. Not. R. Astron. Soc.}  \textbf{410}, 2302 (2011).
\end{thebibliography}
\end{document}